\documentclass[pra,aps,superscriptaddress,groupedaddress,twocolumn]{revtex4}

\usepackage{mathrsfs} 
\usepackage{xspace,amsmath,amsfonts,amsthm,amssymb,amsbsy}
\usepackage{graphicx}
\usepackage{amssymb}
\usepackage{color}
\usepackage{psfrag}
\usepackage{ifsym}
\usepackage{epstopdf}
\usepackage{units} 
\usepackage{bbm}
\usepackage{bbold}
\usepackage[usenames,dvipsnames]{xcolor}
\usepackage[colorlinks=true,citecolor=Cerulean,linkcolor=RubineRed,urlcolor=Cerulean]{hyperref}

\graphicspath{{../images/}}

\usepackage{bbold}

\definecolor{pink}{rgb}{1,0.078,0.57}

\definecolor{green}{rgb}{0,0.7,0.9}

\newcommand{\ket}[1]{| #1 \rangle}
\newcommand{\bra}[1]{\langle #1 |}
\newcommand{\braket}[2] {\langle #1 | #2 \rangle}

\newcommand{\bk}{\mathbf{k}}
\newcommand{\bp}{\mathbf{p}}

\newcommand{\dg}{^{\dagger}}

\newcommand{\ba}{\overrightarrow{\alpha}}

\newcommand{\br}{\mathbf{r}}
\newcommand{\bR}{\mathbf{R}}
\newcommand{\bK}{\mathbf{K}}

\newcommand{\bq}{\mathbf{q}}
\newcommand{\bx}{\mathbf{x}}

\newcommand{\vac}{\ket{v}{}}
\newcommand{\cav}{\bra{v}{}}

\begin{document}

\title{Many-body formalism for thermally excited wave-packets: \\
A way to connect the quantum regime to the classical regime}

\author{Aur\'elia Chenu}
\email[]{achenu@mit.edu}
\affiliation{Massachusetts Institute of Technology, Cambridge, MA 02139, USA}
\affiliation{Singapore-MIT Alliance for Research and Technology, 138602 Singapore}
\author{Monique Combescot}
\affiliation{Institut des NanoSciences de Paris, Universit\'e Pierre et Marie Curie, CNRS, 4 Place Jussieu, 75005 Paris, France}


\begin{abstract}
Free classical particles have well-defined momentum and position, while free quantum particles have well-defined momentum but a position fully delocalized over the sample volume. We develop a many-body formalism based on wave-packet operators that connects these two limits,  the thermal energy being distributed between the state spatial extension and its thermal excitation. The corresponding `mixed quantum-classical' states, which render the Boltzmann operator diagonal, are the physically relevant states when the temperature is finite. The formulation of many-body Hamiltonians in terms of these thermally excited wave-packets and the resulting effective scatterings is provided. 
\end{abstract}

\pacs{}

\maketitle

Connecting wave-packets to quantum particles has been a challenge since the advent of quantum mechanics. 
 Schr\"odinger was aware of the problem   shortly after deriving his equation from de Broglie's matter wave. 
After various infructuous attempts to establish a one-to-one correspondence between wave-packets and free particles, he discovered coherent states of harmonic oscillators, that is,  minimum-uncertainty wave-packets in phase space that follow classical trajectories and remain localized at all times \cite{Schroedinger1926a}. 
How to describe free particles in terms of wave-packets remained a core problem until the understanding of the openness of the system, the seminal work of Joos and Zeh \cite{Joos1985a, SchlosshauerBook}, and the Zurek decoherence program \cite{Zurek2003a}, showing how decoherence leads to a reduced density matrix for the system that represents an improper ensemble of position-space wave-packets whose widths rapidly decrease toward the thermal de Broglie wavelength.

A challenge remains for the representation of thermal equilibrium: Even for the simplest case of free particles, the classical and quantum representations are totally different. Indeed, while the first reads in terms of particles with a well-defined momentum and position, the second involves states with a well-defined momentum but a  position fully delocalized over the sample volume, which are far from any classical-particle state. A representation involving localized quantum states (in the form of wave-packets) is necessary to derive a continuous connection between the quantum and the classical descriptions of a thermal gas. 

We provide such a description here and give a representation of an $N$-particle ideal gas in terms of thermally excited wave-packets that have all the features of classical particles, that is, a well-defined average momentum and position. 
The present derivation provides a more physical picture of this new formalism than the one proposed recently \cite{Chenu2016d}. 
In addition to bridging the gap between traditional quantum and classical representations, the many-body formalism we  develop here provides a versatile basis where the wave-packet spatial extension can be chosen at will. 

Thermal states are ubiquitous in various area of physics: either as the assumed initial state of the system (and thus the starting point for  quantum calculations) or as a state of a reservoir interacting with the system of interest (and thus the central state for describing open systems). 
Due to this central role, studying  different  representations of thermal equilibrium is not only an interesting problem by itself, but also a handy tool.  The versatile formalism we propose allows one to choose the most adequate basis  according to the problem at hand. It can assist in simplifying calculations and facilitating physical interpretation---e.g. the representation of thermal light  in terms of photon number or coherent states leads to significantly different calculation methods, one providing  easier route than the other in many applications. Because our formalism provides  flexibility in  choosing  thermally excited wave packets, it allows for a versatile representation in terms of states that are physically relevant for finite temperature systems. 

The paper is organized as follows. In Sec. \ref{sec:wp}, we first introduce wave-packets and define the creation operator for quantum states having a well-defined momentum and position. 
Section \ref{sec:basis} shows that a complete basis can be constructed from  thermally excited wave-packets. In Section \ref{sec:many-body}, we show how to  use this  basis in a many-body problem. In Sec. \ref{sec:boltzmann}, we determine the wave-packets that diagonalize the Boltzmann operator $e^{-\beta H_0}$, and in Sec. \ref{sec:states}, we discuss the physical relevance  of these states and their link to coherent states for the three-dimensional (3D) harmonic oscillator. In Sec. \ref{sec:discussion}, the practical use of this basis is illustrated by calculating the correlation function through a Green's function procedure. This Section also includes a discussion of the proposed formalism. We then summarize.   
To make this paper easier to read, we have relegated some algebraic derivations to the Appendixes and only kept the ones having some physical insights in the main text. 

\section{wave-packet operators}\label{sec:wp}
Let us consider the  $\ket{\bR,\bK}$ state having a wave function in momentum space reading   
\begin{equation} \label{eq:kwf}
\braket{\bk}{\bR,\bK} = e^{-i \bk\cdot \bR} \braket{\bk-\bK}{\phi}, 
\end{equation}
 where $\ket{\phi}$ characterizes the momentum distribution around $\bK$ of the wave packet at hand. 
By using the $\ket{\bk}$-state closure relation $\sum_\bk \ket{\bk}\bra{\bk} =\mathbb{1}_1$,  we find $\braket{\bR,\bK}{\bR,\bK} = \braket{\phi}{\phi}$
and the  momentum expectation value as 
\begin{equation}\label{eq:k}
\bra{\bR,\bK} \hat{\bk} \ket{\bR,\bK} 
 = \bK \braket{\phi}{\phi} + \sum_\bk (\bk-\bK) \, |\braket{\bk-\bK}{\phi}|^2, 
\end{equation}
the second term reducing to zero for $\ket{\phi}$ chosen such that $\braket{\bk}{\phi} = \braket{-\bk }{\phi}$.  
For such a $\ket{\phi}$, the momentum mean value in the $\ket{\bR,\bK}$ state is  equal to $\bK$,  
\begin{equation}
\bK = \frac{\bra{\bR,\bK}\hat{\bk} \ket{\bR,\bK}}{\braket{\bR,\bK}{\bR,\bK}}.
\end{equation}

If we now turn to $\br$ space and use  $\braket{\br}{\bk} = e^{i \bk \cdot \br} / {L^{D/2}}$, with $\bk$ quantized in $2\pi/L$  for a sample volume $L^D$, 
we find the $\ket{\bR,\bK}$ wave function in $\br$ space as 
\begin{equation}\label{eq:rwf}
\braket{\br}{\bR,\bK}   = \sum_\bk \braket{\br}{\bk}\braket{\bk}{\bR,\bK} = e^{i \bK \cdot (\br - \bR) } \braket{\br-\bR}{\phi}.
\end{equation}
The $\ket{\br}$-state closure relation in a finite volume $L^D$, $\int_{L^D} d\br \: \ket{\br}\bra{\br} = \mathbb{1}_1$, then gives  
\begin{equation}\label{eq:r}
\bra{\bR,\bK} \hat{\br} \ket{\bR,\bK} 
= \bR \braket{\phi}{\phi} + \int_{L^D} d\br (\br-\bR) \, |\braket{\br - \bR}{\phi} |^2,
\end{equation}
the second term reducing to zero for $\braket{\br}{\phi} = \braket{-\br}{\phi}$.  
 The mean value of the $\hat{\br}$ operator in the $\ket{\bR,\bK}$ state is then equal to $\bR$,   
\begin{equation}
\bR = \frac{ \bra{\bR,\bK} \hat{\br} \ket{\bR,\bK} } {\braket{\bR,\bK}{\bR,\bK}}.
\end{equation}

All this shows that, for a symmetrical $\ket{\phi}$ the operator $a\dg_{\bR,\bK}$, defined as $\ket{\bR,\bK} = a\dg_{\bR,\bK} \vac$ where $\vac$ denotes the vacuum state, creates a wave-packet with average position $\bR$ and average momentum $\bK$.
The state $\ket{\phi}$ characterizes the wave-packet extension, either in $\br$ space around $\bR$ through $\braket{\br-\bR}{\phi}$, or in $\bk$ space around $\bK$ through $\braket{\bk-\bK}{\phi}$. 
From the closure relations of the $\ket{\bk}=a\dg_\bk \vac$ states, and $\ket{\br}=a\dg_\br \vac$ states, we find that this creation operator expands as  
\begin{equation} \label{eq:wp_def}
a\dg_{\bR,\bK} = \sum_\bk a\dg_\bk\braket{\bk}{\bR,\bK}= \int_{L^D} d\br \: a\dg_\br \braket{\br}{\bR,\bK} ,
\end{equation}
with $\braket{\bk}{\bR,\bK}$ and $\braket{\br}{\bR,\bK}$ given in Eqs.~(\ref{eq:kwf},\ref{eq:rwf}).  
The $\ket{\phi}$ extension controls the number of $\ket{\br}$ and $\ket{\bk}$ states making the $\ket{\bR,\bK}$ wave packet.

For a highly peaked function in $\bk$ space, that is,  $\braket{\bk}{\phi} = \delta_{\bk \mathbf{0}}$, we find 
 $\braket{\br}{\phi}= \sum_\bk \braket{\br}{\bk} \braket{\bk}{\phi} = L^{-D/2}$; so,   $\ket{\phi}$ is flat in $\br$ space. 
The $a\dg_{\bR,\bK}$ operator then reduces to 
\begin{equation}\label{eq:peaked-k}
 a\dg_\bK e^{- i \bK \cdot \bR} = \int_{L^D} d\br' a\dg_{\br'+\bR } \braket{\br'}{\bK}.
\end{equation}
So, for  such a $\ket{\phi}$, the extension of  $a\dg_{\bR,\bK}$  in terms of  $a\dg_\bk$  is highly peaked on $a\dg_\bK$ but fully delocalized in terms of $a\dg_\br$; and vice-versa for $\ket{\phi}$ highly peaked in $\br$ space.

\section{Complete basis made of wave-packets \label{sec:basis}}
While the $\ket{\bk}$ states form an orthogonal set, i.e., $\braket{\bk}{\bk'} = \delta_{\bk \bk'}$, the $\ket{\bR,\bK}$ states do not, due to their spatial extension. Indeed, we find from Eq.~(\ref{eq:kwf})    
\begin{eqnarray} \label{eq:overlap}
\braket{\bR',\bK'}{\bR,\bK} 
=\braket{\phi}{\phi} e^{i \frac{\bK'+ \bK}{2} \cdot (\bR' - \bR)} \delta_\phi(\bR'-\bR, \bK'-\bK),
\end{eqnarray}
where the function $\delta_\phi(\bR'-\bR, \bK'-\bK)$,  given by 
\begin{align}\label{eq:delta_g}
&\delta_\phi(\bR'-\bR, \bK'-\bK) = \braket{\phi}{\phi}^{-1}   \\
& \qquad  \times \sum_\bp e^{i \bp \cdot (\bR'-\bR)} \braket{\phi}{\bp - \frac{\bK'-\bK}{2}} \braket{\bp+ \frac{\bK'-\bK}{2}}{\phi}, \nonumber
\end{align}
characterizes the wave-packet overlap. It is such that  $\delta_\phi(\mathbf{0},\mathbf{0}) = 1$ and  $\delta_\phi ( \bR,\bK) \approx 0$ for $\bK$ or $\bR$ larger than the $\phi$ extension in the respective space.  
In the case of a highly peaked $\ket{\phi}$ in $\bk$ space, that is, $\braket{\bk}{\phi} = \delta_{\bk \mathbf{0}}$, the $\delta_\phi(\bR'-\bR, \bK' - \bK)$ function reduces to $\delta_{\bK' \bK}$ and $\braket{\bR,\bK}{\bR',\bK'}$ reduces to $\delta_{\bK' \bK} e^{i \bK \cdot (\bR'-\bR)} \braket{\phi}{\phi}$. 

Although nonorthogonal, the $\ket{\bR,\bK}$ states still form a complete basis. 
Indeed, for $\braket{\phi}{\phi}=1$ as taken for simplicity in order to have normalized states, namely,  $\braket{\bR,\bK}{\bR,\bK} = 1$, the operator 
\begin{equation} \label{eq:id}
\sum_\bK \int_{L^D} \frac{d\bR}{L^D} \ket{\bR,\bK}\bra{\bR,\bK} =  \mathbb{1}_1
\end{equation}
is the identity operator in the one-particle subspace. 
The above equation allows us to write the $a\dg_\bk$ operator in terms of the normalized $a\dg_{\bR,\bK}$ operators  as 
\begin{equation}\label{eq:ak}
a\dg_\bk = \sum_\bK \int_{L^D} \frac{d\bR}{L^D} a\dg_{\bR,\bK} \braket{\bR,\bK}{\bk} 
\end{equation}
and similarly for $a\dg_\br$, 
\begin{equation}
a\dg_\br = \sum_\bK \int_{L^D} \frac{d\bR}{L^D} a\dg_{\bR,\bK} \braket{\bR,\bK}{\br}, 
\end{equation}
the $\braket{\bR,\bK}{\bk}$ and $\braket{\bR,\bK}{\br}$ prefactors being given respectively  by Eqs. (\ref{eq:kwf}) and (\ref{eq:rwf}). 

More generally, the closure relation in the $N$-particle subspace 
\begin{equation}
\mathbb{1}_N = \frac{1}{N!} \sum_{\{ \bk_i \} } a\dg_{\bk_1} \dots a\dg_{\bk_N} \vac \cav a_{\bk_N} \dots a_{\bk_1}
\end{equation}
takes a similar compact form in terms of  $a\dg_{\bR,\bK}$ operators,  
\begin{equation} \label{eq:id}
\begin{split}
\mathbb{1}_N = \frac{1}{N!} \sum_{ \{ \bK_i \}} \int_{L^D} \left\{ \frac{d\bR_i}{L^D} \right\} & a\dg_{\bR_1, \bK_1} \dots a\dg_{\bR_N, \bK_N} \vac  \\ 
&\cav a_{\bR_N, \bK_N} \dots a_{\bR_1, \bK_1}.
\end{split}
\end{equation}
Thus, the $a\dg_{\bR_1,\bK_1}\dots a\dg_{\bR_N,\bK_N}\vac$ states form a complete basis for the $N$-particle subspace and as such   can be used to decompose any state or write any many-body operator.

\section{Many-body Hamiltonians \label{sec:many-body}}
We now construct a many-body formalism in terms of the normalized wave-packet operators $a\dg_{\bR,\bK}$ defined above. 
Using Eq.~(\ref{eq:ak}), we first note that the particle-number operator remains diagonal,  
\begin{equation}
\hat{N} = \sum_\bk a\dg_\bk a_\bk =  \sum_\bK \int_{L^D} \frac{d \bR}{L^D} a\dg_{\bR,\bK} a^{ }_{\bR,\bK}.
\end{equation}

If we now consider the free Hamiltonian, Eq. (\ref{eq:ak}) leads to 
\begin{eqnarray}\label{eq:H0}
H_0 &=& \sum_\bk \epsilon_\bk a\dg_\bk a_\bk \\ 
&=& \sum_{\bK' \bK} \int_{L^D} \frac{d\bR'}{L^D}  \frac{d \bR}{L^D} a\dg_{\bR',\bK'} a^{}_{\bR,\bK} \bra{\bR', \bK'} H_0 \ket{\bR,\bK} \nonumber,
\end{eqnarray}
with the prefactor given by 
\begin{align} \label{eq:H0}
&\bra{\bR',\bK'} H_0 \ket{\bR,\bK}   
= e^{i \frac{\bK' + \bK}{2} \cdot (\bR' - \bR)}   \times  \\
&  \sum_\bp  \epsilon_{\bp + \frac{\bK'+\bK}{2}} e^{i \bp \cdot (\bR' - \bR) } \braket{\phi}{\bp - \frac{\bK' -\bK}{2}} \braket{\bp + \frac{\bK'-\bK}{2}}{\phi}. \nonumber
\end{align}

In the case of a highly peaked  function in momentum space, that is, $\braket{\bk}{\phi} = \delta_{\bk \mathbf{0}}$, the $\phi$ part of the above equation reduces to $\delta_{\bK' \bK} \delta_{\bp \mathbf{0}}$; so $H_0$ reduces to $\sum_\bK \epsilon_\bK a\dg_\bK a^{}_\bK$, as expected (see Appendix A). 

In the same way, the two-particle potential of a many-body Hamiltonian 
\begin{equation} \label{eq:V}
V = \frac{1}{2} \sum_ {\bq} V_{\bq} \sum_{\bk_1 \bk_2} a\dg_{\bk_1 + \bq} a\dg_{\bk_2 - \bq} a^{}_{\bk_2} a^{}_{\bk_1}
\end{equation}
reads, with the help of Eq. (\ref{eq:ak}), 
\begin{align}
V =\sum_{\{ \bK  \}} \int_{L^D} \left\{ \frac{ d\bR}{L^D} \right\}  &
\: \mathcal{V}\left( 
\begin{matrix}
\bR_2' \bK_2' & \bR_2 \bK_2 \\
\bR_1'\, \bK_1' & \bR_1 \bK_1 \\
\end{matrix}\right) \\
& \times 
 a\dg_{\bR_1',\bK_1'} a\dg_{\bR'_2,\bK'_2}  a^{}_{\bR_2,\bK_2} a^{}_{\bR_1,\bK_1}, \nonumber
\end{align}
where the scattering amplitude between wave-packets splits as 
\begin{equation}
\begin{split}
\mathcal{V}\left( \begin{matrix}
\bR_2' \bK_2' & \bR_2 \bK_2 \\
\bR_1' \,\bK_1' & \bR_1 \bK_1 \\
\end{matrix}\right) = \sum_{\bq} & V_\bq  u_\bq(\bR_1',\bK_1';\bR_1,\bK_1) \\
\times &u_{-\bq}(\bR'_2,\bK'_2; \bR_2,\bK_2),
\end{split}
\end{equation}
 the $\bq$-channel amplitude being given by   
\begin{equation}
\begin{split}
u_{\bq}(\bR',\bK';\bR,\bK) =& \sum_\bk \braket{\bR',\bK'}{\bk+ \bq} \braket{\bk}{\bR,\bK} \\
=& e^{i \bq \cdot \bR'} \braket{\bR',\bK'-\bq}{\bR,\bK}.
\end{split}
\end{equation}
The $\delta_\phi$ function that appears in the above scalar product, [see  Eq.~(\ref{eq:overlap})], forces $\bK'$ to be close to $\bK + \bq$ and $\bR'$ to be close to $\bR$. 
For a very highly peaked function in momentum space, $u_\bq(\bR',\bK'; \bR,\bK)$ reduces to $\delta_{\bK' \bK} e^{i (\bK + \bq)\cdot \bR'} e^{-i \bK \cdot\bR} $ and we recover the potential given in Eq. (\ref{eq:V}), as expected. 
 For a broad function, the scattering potential is broadened in space and momentum.

\section{Boltzmann operator}\label{sec:boltzmann}
Let us now consider the Boltzmann operator for a free system, namely, $e^{-\beta H_0}$ with $\beta = (k_B T)^{-1}$,  that describes thermal equilibrium.  

(i) As  $\left[H_0 \, , a\dg_{\bk} \right]= \epsilon_\bk a\dg_{\bk} $ yields  
$e^{-\beta H_0} a\dg_\bk = a\dg_{\bk} e^{-\beta (H_0 + \epsilon_\bk)}$, the Boltzmann operator in the $N$-particle subspace takes a diagonal form when written with the help of the $\ket{\bk}$-state closure relation. 
Indeed, since $e^{-\beta H_0} \vac = \vac$, we  readily find 
\begin{eqnarray} \label{eq:th_ak}
\big\{e^{-\beta H_0}\big\}_N {=} e^{-\beta H_0} \frac{1}{N!} \sum_{\{\bk_i\}} a\dg_{\bk_1} \dots a\dg_{\bk_N} \vac \cav a_{\bk_N} \dots a_{\bk_1} \: \: \\
= \frac{1}{N!} \sum_{\{\bk_i\}}
\Big(e^{-\beta \sum_{i=1}^N\epsilon_{\bk_i} }\Big) 
 a\dg_{\bk_1} \dots a\dg_{\bk_N} \vac  \cav a_{\bk_N} \dots a_{\bk_1}.\:  \nonumber
\end{eqnarray}
The $a\dg_\bk$ operators used in this representation do not depend on the temperature $T$; all the thermal energy is carried in the Boltzmann factors $e^{-\beta \epsilon_\bk}$, which physically correspond to the probability for the $\ket{\bk}=a\dg_\bk \vac$ state to be thermally occupied. The $\ket{\bk}$ state eigenenergy is equal to $\epsilon_\bk$, and the $\ket{\bk}$ state energy variance $\sigma_{\bk} = \bra{\bk} H_0^2 \ket{\bk}-\bra{\bk} H_0 \ket{\bk}^2$ is equal to zero.

(ii)  
We can also write the Boltzmann operator using the $\ket{\br}$-state closure relation, that is 
 \begin{equation}\label{eq:H0inr}
 \begin{split}
\big\{e^{{-}\beta H_0}\big\}_N {=}&
 e^{-\beta H_0} \frac{1}{N!} \int_{L^D} d{\{\br_i \}}  \\ 
 & \times a\dg_{{\br_1}} {\dots} a\dg_{\br_N} \vac \cav a_{\br_N} {\dots} a_{{\br_1}}. 
 \end{split}
\end{equation}
To pass $e^{-\beta H_0}$ over $a\dg_{\br}$, we use the relations between the $a\dg_\br$ and $a\dg_\bk$ operators, namely  
\begin{eqnarray}\label{eq:commutator}
a\dg_\br = \sum_\bk a\dg_\bk \braket{\bk}{\br}, \quad \quad
a\dg_\bk = \int_{L^D}d\br \ a\dg_\br \braket{\br}{\bk} \label{eq:ak2}. 
\end{eqnarray}
They readily give
\begin{eqnarray}\label{eq:H0ar}
e^{-\beta H_0 } a\dg_{\br_1} &=& e^{-\beta H_0} \sum_{\bk} a\dg_{\bk} \braket{\bk}{{\br_1}} = \sum_\bk a\dg_\bk \braket{\bk}{{\br_1}} e^{-\beta (H_0 + \epsilon_\bk)} \nonumber \\
&=& \Bigg( \int_{L^D} d{\br_1'} a\dg_{{\br_1'}} \bra{{\br_1'}} e^{-\beta H_0} \ket{{\br_1}} \Bigg) e^{-\beta H_0}.
\end{eqnarray}

To go further and write Eq.~(\ref{eq:H0inr}) in a diagonal form, we must decouple ${\br'_1}$ from ${{\br_1}}$ in the above equation.
This is done 
 by splitting $e^{-\beta H_0}$ as $e^{-\beta H_0/2}e^{-\beta H_0/2}$ and by inserting the $\ket{\br}$-state closure relation between the two $e^{-\beta H_0/2}$ operators. 
 This yields 
 \begin{align}\label{8}
& \bra{{\br_1'}} e^{-\beta H_0} \ket{{\br_1}} = \bra{{\br_1'}}e^{{-}\beta H_0/2} \\
&\times \left(\int_{L^D} d\bR_1 \ket{{\br_1}+{\br_1'}-\bR_1 }\bra{{\br_1} + {\br_1'}- \bR_1}\right) e^{{-}\beta H_0/2}\ket{{\br_1}}. \nonumber
\end{align}
 Due to the translational invariance of the system,  $\bra{{\br_1'}} e^{-\beta H_0} \ket{{\br_1}}$
 depends on $|{\br_1'} - {\br_1}|$ only, as directly seen from 
\begin{equation} \label{eq:prefactor}
\bra{{\br_1'}} e^{-\beta H_0} \ket{{\br_1}} =  \sum_\bk  \braket{{\br_1'}}{\bk}  e^{-\beta \epsilon_\bk}      \braket{\bk}{{\br_1}}    = \frac{Z_T}{L^D}e^{- |{\br_1'} - {\br_1}|^2/{\lambda}_T^2}
\end{equation}
where $\lambda_T$ is the thermal length that we defined as
\begin{equation}
\beta^{-1} = k_B T = 4\  \hbar^2 / (2m {\lambda}^2_T),
\end{equation}
and $Z_T= \sum_{\bk} e^{-\beta \epsilon_\bk} =(L/{\lambda}_T\sqrt{\pi})^D$ is the partition function for one free particle.
This shows that $ \bra{\br+\br'-\bR} e^{-\beta H_0/2}\ket{\br}$ reduces to $\braket{\br'-\bR}{\phi_T} $ with 
\begin{equation} \label{eq:phiT}
\ket{\phi_T}=e^{-\beta H_0/2} \ket{\br = \bold{0}}.
\end{equation}

Equations (\ref{eq:H0ar},\ref{8}) then give 
\begin{equation}\label{7'}
e^{-\beta H_0 } a\dg_{\br_1} = \left( \int_{L^D} d\bR_1 \ a\dg_{\bR_1,T} \braket{\phi_T}{\br_1-\bR_1} \right) e^{-\beta H_0}, 
\end{equation}
with  $a\dg_{\bR_1,T}$  defined as 
\begin{eqnarray}\label{eq:aRT}
a\dg_{\bR_1,T} &=& \int_{L^D} d\br_1' \ a\dg_{\br_1'} \braket{\br_1'-\bR_1}{\phi_T} \\
&=& \sum_\bk a\dg_\bk \: e^{- i \bk \cdot \bR} \braket{\bk}{\phi_T}, \nonumber
\end{eqnarray}
which follows from Eq.~(\ref{eq:commutator}).
By noting that the $\braket{\phi_T}{\br_1-\bR_1}$ factor in Eq. (\ref{7'}) can be used in Eq.~(\ref{eq:H0inr}) to produce the $a^{ }_{\bR_1 ,T}$ operator, 
it becomes straightforward to show that the Boltzmann operator in the $N$-quantum particle subspace written with the help of the $\ket{\br}$-state closure relation as in Eq. (\ref{eq:H0inr})  
also takes a diagonal form
 \begin{eqnarray} \label{eq:th_aRT}
\big\{e^{-\beta H_0}\big\}_N &{=}& \frac{1}{N!} \int_{L^D} d{\{\bR_i\}} \\
&&\times a\dg_{\bR_1,T}{ \dots} a\dg_{\bR_N,T} \vac \cav a_{\bR_N,T} {\dots} a_{\bR_1,T}. \nonumber
\end{eqnarray}
In the next section, we will study the physics associated with the $a\dg_{\bR,T}$ operator, and show that it creates a wave-packet with spatial extension $\lambda_T$ around $\bR$, and a momentum equal to zero. 

(iii) It also is possible to write $e^{-\beta H_0}$ in a diagonal form in terms of operators that create wave-packets with non-zero  momentum, by splitting the thermal energy $k_B T$,  as first introduced by Hornberger and Sipe \cite{Hornberger2003a}, into a part accounting for the spatial extension of the wave-packet and a part accounting for its kinetic energy, namely 
\begin{equation}\label{eq:splitting}
 k_B T = k_B T_\mathcal{R} + k_B T_\mathcal{K}   .
\end{equation}
Using Eq.~(\ref{eq:prefactor}), it is then easy to check that $\bra{\br_1'} e^{-\beta H_0} \ket{\br_1}$ splits as
\begin{eqnarray}
\bra{\br_1'} e^{-\beta H_0} \ket{\br_1}&=&
\Big(\frac{\sqrt{\pi}\lambda_{T_\mathcal{R}}\lambda_{T_\mathcal{K}}}{\lambda_{T}}\Big)^{D} \\
&&\times \bra{\br_1'} e^{-\beta_\mathcal{R} H_0} \ket{\br_1}\bra{\br_1'} e^{-\beta_\mathcal{K} H_0} \ket{\br_1}. \nonumber
\end{eqnarray}

 To go further, we again have to decouple $\br'_1$ from $\br_1$. In the $\bra{\br_1'} e^{-\beta_\mathcal{R} H_0} \ket{\br_1}$ part, we use the $\ket{\br}$-state closure relation, as done to get Eq.~(\ref{8}), 
  and we obtain Eq.~(\ref{eq:prefactor}) with $\beta$ replaced by $\beta_\mathcal{R}$. In the $\bra{\br_1'} e^{-\beta_\mathcal{K} H_0} \ket{\br_1}$ part of the above equation, we use the $\ket{\bk}$-state closure relation. This yields 
 \begin{equation}
\bra{\br'_1} e^{-\beta_\mathcal{K} H_0} \ket{\br_1} = \sum_{\bK_1}  \braket{\br_1' -\bR_1}{\bK_1}e^{-\beta_\mathcal{K} \epsilon_{\bK_1}} \braket{\bK_1}{\br_1 -\bR_1}. 
\end{equation}
The  procedure we have used  to obtain Eq.~(\ref{7'}) allows us to rewrite Eq.~(\ref{eq:H0ar}) as 
\begin{align} \label{eq:mix_th}
& e^{-\beta H_0}a\dg_{\br_1} = \Big(\frac{\sqrt{\pi} \lambda_{T_\mathcal{R}}\lambda_{T_\mathcal{K}}}{\lambda_{T}}\Big)^{D} 
\sum_{\bK_1} e^{-\beta_\mathcal{K}  \epsilon_{\bK_1} } \times    \qquad \\ 
& \left(\int_{L^D} d{\bR_1} \: a\dg_{\bR_1,\bK_1,T_\mathcal{R}}  \braket{\phi_{T_\mathcal{R}}}{\br_1-\bR_1} \braket{\bK_1}{\br_1-\bR_1} \right) 
e^{-\beta H_0}, \nonumber
\end{align}
the $a\dg_{\bR_1, \bK_1,T}$  operator being defined as 
\begin{eqnarray}\label{eq:aRKT}
a\dg_{\bR_1,\bK_1,T} &=& L^{-D/2} \int_{L^D} d\br_1' \: a\dg_{ \br_1'} e^{i \bK_1 \cdot (\br'_1 - \bR_1)} \braket{\br_1'-\bR_1}{\phi_{T}} \nonumber \\
&=& L^{-D/2} \sum_{\bk} a\dg_\bk \: e^{-i \bk \cdot \bR_1} \braket{\bk-\bK_1}{\phi_T},
\end{eqnarray}
with $\ket{\phi_T}$ defined in Eq. (\ref{eq:phiT}).
The above equations are similar to Eqs.~(\ref{eq:kwf},\ref{eq:rwf}) with $\ket{\phi}$ replaced by $L^{-D/2} \ket{\phi_T}$. 

It is then straightforward to show that Eq.~(\ref{eq:H0inr}) also reads
\begin{align} \label{eq:th_aRKT}
\big\{e^{-\beta H_0}\big\}_N {=}&\frac{1}{N!}
\Big(\frac{\sqrt{\pi}\lambda_{T_\mathcal{R}}\lambda_{T_\mathcal{K}}}{\lambda_{T}}\Big)^{ND}  \sum_{\{\bK_i\}}
\Big(e^{-\beta_\mathcal{K} \sum_{i=1}^N\epsilon_{\bK_i} }\Big)  \nonumber \\ 
\times \int_{L^D} d{\{\bR_i\}}&\: a\dg_{\bR_1,\bK_1,T_\mathcal{R}}{ \dots} \, a\dg_{\bR_N,\bK_N,T_\mathcal{R}} \vac \nonumber\\
& \cav a_{\bR_N,\bK_N,T_\mathcal{R}} {\dots} \, a_{\bR_1,\bK_1,T_\mathcal{R}},
\end{align}
which is diagonal in the basis formed by the states $a\dg_{\bR_1,\bK_1,T_\mathcal{R}}{ \dots} \, a\dg_{\bR_N,\bK_N,T_\mathcal{R}} \vac$. The physics associated with the  $a\dg_{\bR,\bK, T}$ operator is discussed in the next section. We will show that this operator creates a wave-packet with average momentum $\bK$ and average position $\bR$, the  position  extension being controlled by $\lambda_T$.

\section{Thermally-excited wave-packets}\label{sec:states}
\subsection{Properties of the $\ket{\bR, T} = a\dg_{\bR,T}\vac$ state} \label{sec:static}

Let us first understand the physics associated with the $a\dg_{\bR, T}$ operator making $e^{-\beta H_0}$ diagonal.

(i) As seen from Eqs.~(\ref{eq:prefactor}, \ref{eq:phiT}), the $\braket{\br}{\phi_T}$ wave function 
 is localized at $\br = \bold{0}$ with a spatial extension ${\lambda}_T/\sqrt{2}$; so, the operator $a\dg_{\bR, T}$ defined in Eq.~(\ref{eq:aRT}) creates a wave-packet $\ket{\bR, T} = a\dg_{\bR,T}\vac$ with wave function $\braket{\br}{\bR,T} = \braket{\br-\bR}{\phi_T}$ localized around $\bR$ with a spatial extension also scaling as $\lambda_T$, the norm of  the $\ket{\bR,T}$ state being given by  
 \begin{equation}
 \braket{\bR,T}{\bR,T} = \braket{\phi_T}{\phi_T} = \left( \frac{1}{\lambda_T \sqrt{\pi}} \right)^D.
 \end{equation}
 
\begin{figure}
\includegraphics[width = 1\columnwidth]{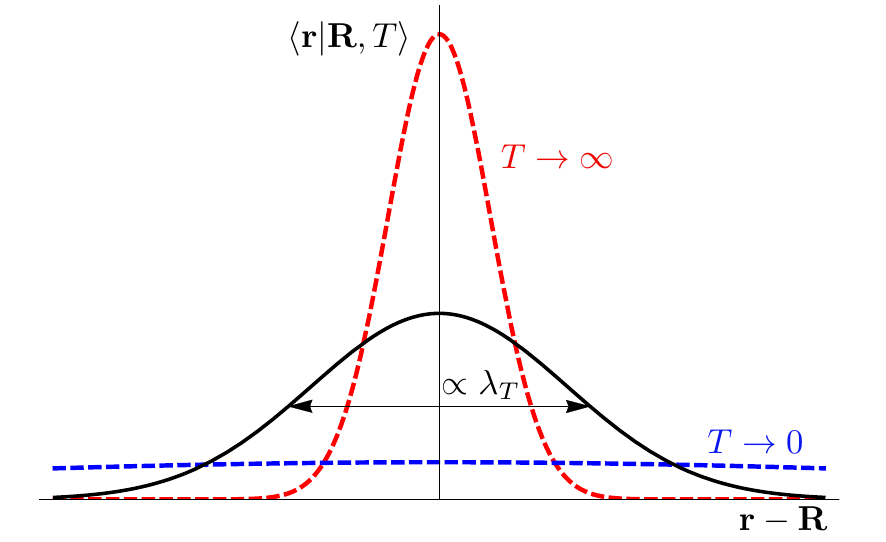}
\caption{Wave function of the state $\ket{\bR,T}=a\dg_{\bR,T}\vac$, as defined in Eq.~(\ref{eq:aRT}),  represented in $\br$ space for various temperatures. The higher the temperature, the more localized the state.  \label{fig:phi}}
\end{figure}

 The scalar product of two $\ket{\bR,T}$ states having the same $T$ is given by  
\begin{equation}
\begin{split}
\braket{\bR,T}{\bR',T}  =& \int_{L^D} d\br \braket{\phi_T}{\br -\bR } \braket{\br - \bR'}{\phi_T} \\ 
 \equiv&\braket{\phi_T}{\phi_T} \: \delta_{\phi_T}(\bR - \bR'), 
 \end{split}
\end{equation}
 with $\delta_{\phi_T}(\bR)$ equal to 1 for $\bR=0$ and to $\sim0$ for $|\bR|$ large compared to $\lambda_T$.
  Hence, due to the wave-packet spatial extension, two ${a}\dg_{\bR,T}\vac$ states with different $\bR$'s are not  orthogonal. 
  As illustrated in Fig. \ref{fig:phi}, when $T$ goes to zero, $\lambda_T$ goes to infinity and the wave-packet is fully delocalized in space: It then looks like a free quantum particle. By contrast, when $T$ goes to infinity, $\lambda_T$ goes to zero and the wave-packet is fully localized in space: It then looks like a classical particle.

 (ii) 
 This wave-packet does not move as seen from the mean value of the momentum operator $\hat{\bk}$,  which reduces to zero for $\ket{\phi_T}$ being a symmetrical state, $\braket{\bk}{\phi_T} = \braket{-\bk}{\phi_T}$ (see Appendix B).

(iii)  The energy of the $\ket{\bR, T}$ state reads (see Appendix) 
\begin{equation} \label{eq:H0_aRT}
\frac{\bra{\bR, T} \: H_0 \: \ket{\bR, T} }{\braket{\bR, T}{\bR, T}}  =  \frac{D}{2} k_B T, 
\end{equation}
which is exactly equal to the energy of a classical particle when the temperature is $T$: Within the $a\dg_{\bR, T}$ representation, the thermal energy entirely lies in the spread of the wave-packet operators.
 
(iv) The energy variance  $\sigma_{\bR,T}$ of the $a\dg_{\bR, T}\vac$ state, found equal to $\frac{D}{2} (k_B T)^2$ (see Appendix B), 
is that of a classical particle. 

\subsection{Properties of the $\ket{\bR,\bK,T} = a\dg_{\bR,\bK,T} \vac$ state}
\begin{figure*}
\includegraphics[width = 1\textwidth]{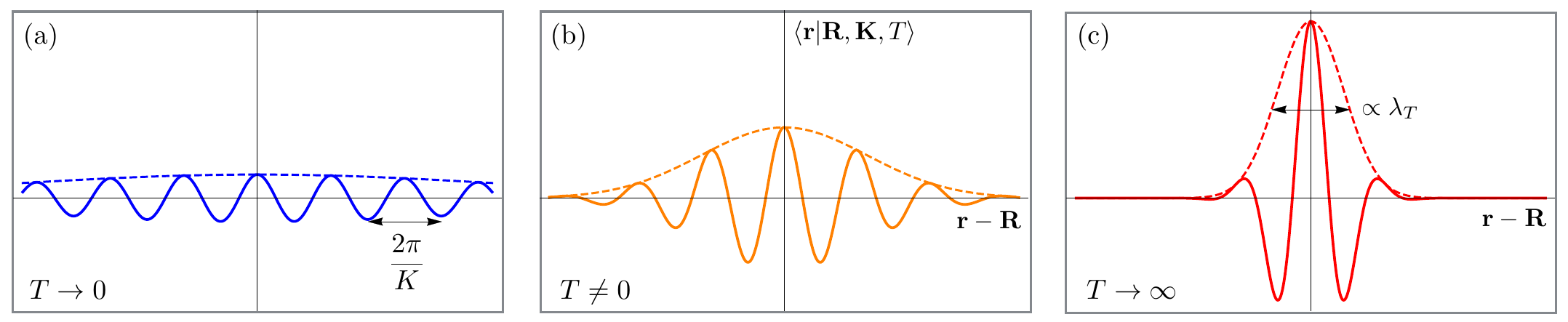}
\caption{ Real part (solid lines) and absolute value (dashed lines) 
 of the coordinate representation of the wave-packet state $\ket{\bR,\bK,T} \equiv a\dg_{\bR,\bK,T} \vac$  for increasing temperatures [from (a) to (c)]. When $T\to0$, the $\ket{\bR,\bK,T}$ state looks like the free state $\ket{\bK}$. For non-zero temperature, the thermal energy $k_B T$ is distributed between the state spatial extension and its thermal excitation, as evidenced by the state average momentum $\bK$.  \label{fig:wp}}
\end{figure*}

(i) The operator $a\dg_{\bR,\bK,T}$ defined in Eq.~(\ref{eq:aRKT})  is similar to the wave-packet operator $a\dg_{\bR,\bK}$ defined in Section \ref{sec:wp}, with $\ket{\phi}$ replaced by $L^{-D/2} \ket{\phi_T}$. 
So it creates a wave-packet localized around $\bR$ with a momentum mean value equal to $\bK$, its spatial extension scaling as  $\lambda_{T}$.
 From (\ref{eq:overlap}), the scalar product of two same-$T$ states  is equal to
   \begin{eqnarray}
\braket{\bR',\bK', T}{\bR,\bK, T} 
=&L^{-D} \braket{\phi_T}{\phi_T} e^{i\frac{\bK'+\bK}{2} .(\bR'-\bR)} \nonumber \\ 
&\times 
\delta_{\phi_T}(\bR' - \bR,\bK'-\bK),
\end{eqnarray}
where the extension of  $\delta_{\phi_T}(\bR,\bK)$, defined in Eq. (\ref{eq:delta_g}), is now controlled by the temperature $T$. 
As illustrated in Fig. \ref{fig:wp}, when $T$ goes to zero,  $\lambda_T$ goes to infinity and  the $\ket{\bR ,\bK,T}$ wave-packet is fully delocalized in space: It then looks like a free quantum particle. By contrast, when $T$ goes to infinity, $\lambda_T$ goes to zero and the wave-packet is fully localized in space: It then looks like a classical particle.
We can also note that the above overlap is  $\sim0$ for $|\bR|\gg\lambda_T$ or $|\bK|\gg1/\lambda_T$; 
so $\lambda_T$ corresponds to the de Broglie wavelength characterizing the spatial extension of the wave-packet.  

(ii) It is possible to show that the $\ket{\bR,\bK,T}$ wave-packet moves with an average momentum that stays equal to $ \bK$,  despite the wave-packet spatial spread, 
the $\bK$-momentum probability at equilibrium  being controlled by the Boltzmann factor $e^{-\beta_\mathcal{K} \epsilon_\bK}$, as seen from Eq.~(\ref{eq:th_aRKT}).

 We can also calculate the uncertainties in momentum and position. As $ \bra{\bR,\bK, T } \hat{k}_x^2 \ket{\bR,\bK, T} /\braket{\bR,\bK, T}{\bR,\bK, T}  = K_x^2 + 2 / \lambda_T^2$  for each Cartesian coordinate, 
 while $ \bra{\bR,\bK,T} \hat{x} \ket{\bR,\bK,T} / \braket{\bR,\bK, T}{\bR,\bK, T} = \bR_x$
 and $\bra{\bR, \bK ,T} \hat{x}^2 \ket{\bR, \bK ,T} /\braket{\bR,\bK, T}{\bR,\bK, T} = \bR_x^2 + \lambda_T^2 / 8$, 
 the momentum uncertainty of the  $\ket{\bR, \bK, T}$ state is equal to  
  \begin{equation*}
  \begin{split}
\left(\Delta k_x \right)^2 &\equiv \frac{\bra{\bR,\bK, T} \hat{k}_x^2 \ket {\bR, \bK,T}}{\braket{\bR,\bK,T}{\bR,\bK,T}} - \frac{\bra{\bR,\bK,T} \hat{k}_x \ket {\bR,\bK,T}^2}{\braket{\bR,\bK,T}{\bR,\bK,T}^2}\\
&=\frac{2}{\lambda_T^2}
\end{split}
\end{equation*}
 and the position uncertainty is equal to $\left(\Delta x \right)^2={\lambda_T^2}/{8}$.
Not surprisingly, the fluctuations around the mean position and momentum depend on the wave-packet extension. These uncertainties lead to $ \Delta {k_x}   \Delta x = 1 / 2$, which shows that  the $\ket{\bR,\bK,T}$ states correspond to minimum uncertainty states, independently from their extension.

(iii) The energy  mean value of the $\ket{\bR, \bK, T}$ state is equal to 
\begin{equation}
\frac{{\bra{\bR, \bK, T} H_0 \ket{\bR, \bK, T}} }{\braket{\bR,\bK,T}{\bR,\bK,T}}= \frac{D}{2} k_B T + \epsilon_\bK. 
\end{equation}
This energy has a classical component given by the first term and a quantum component given by the second term. 
When $T \to 0$, the wave-packet energy tends to the energy of the quantum state $\ket{\bK} $; by contrast, for $\bK =0$, we  recover the energy of the zero average-momentum state $\ket{\bR,T}$, given in Eq.~(\ref{eq:H0_aRT}), which is totally classical.

(iv) Similarly, the variance in energy of the $\ket{\bR,\bK,T}$ state is given by (see Appendix B)
\begin{eqnarray} \label{eq:sigma_wp}
\sigma_{\bR, \bK, T} &=&\frac{1}{Z_T} \sum_{\bk} \epsilon_{\bk+\bK}^2 e^{-\beta \epsilon_{\bk}}   - \left( \frac{D}{2} k_B T + \epsilon_\bK \right)^2,  
\end{eqnarray}
which again has a classical part and a thermally excited quantum part.

(v) The time evolution of the $\ket{\bR,\bK,T}$ state follows from Eq.~(\ref{eq:aRKT}) as 
\begin{eqnarray}\label{eq:evolution}
\ket{\bR,\bK,T}_t &=& e^{-i H_0 t} \ket{\bR,\bK,T} \\
&=&L^{-D/2} \sum_\bk \ket{\bk} e^{-i ( \bk\cdot \bR + \epsilon_\bk t) } \braket{\bk-\bK}{\phi_T}. \nonumber
\end{eqnarray}
This readily shows that the norm of this state stays constant and equal to $\braket{\phi_T}{\phi_T} / L^D $. 
 The wave-packet average momentum also stays  constant and equal to $\bK$, as expected for a non-interacting system having a   Hamiltonian equal to $H_0$. 
If we now consider the time evolution of the average position,  we find (see Appendix)
\begin{equation}\label{eq:Rt}
{_t \bra{\bR,\bK,T} \hat{\br}\ket{\bR,\bK,T}_t}  \approx \frac{ \braket{\phi_T}{\phi_T}}{L^D}  \bR_t ,
\end{equation}
with $\bR_t  \equiv \bR + \bK t / m$: 
The wave-packet  moves with a velocity $\bK/m$, as a classical particle, without changing its velocity or its symmetrical shape, as can be  directly seen from 
\begin{eqnarray} \label{eq:RKTt}
\braket{\br}{\bR,\bK,T}_t &\approx& L^{-D/2} e^{i \epsilon_\bK t} e^{i \bK \cdot(\br - \bR_t)} \braket{\br - \bR_t }{\phi_T}  \nonumber \\
&= & e^{i \epsilon_\bK t} \braket{\br}{\bR_t,\bK}, 
\end{eqnarray}
with $\braket{\br}{\bR,\bK}$ given in Eq.~(\ref{eq:rwf}) for $\ket{\phi} = L^{-D/2} \ket{\phi_T}$. 
 We could expect the wave-packet to spread over time because each $\bk$ component travels at its own velocity, but  the approximation   $\epsilon_\bk\approx \epsilon_\bK + (\bk - \bK)\cdot \bK/m$ that we used to obtain Eq.~(\ref{eq:Rt}) amounts to neglecting these other components.

(vi) 
The $\ket{\bR,\bK,T}$ states can be interpreted as a form of coherent states. 
To recognize this, let us consider the Hamiltonian of the 3D harmonic oscillator 
\begin{equation}
\begin{split}
H =& \frac{1}{2m} \hat{\bp}^2 + \frac{1}{2} m \omega^2 \hat{\br}^2 = h_{\bf x} + h_{\bf y } + h_{\bf z}, \\
\end{split}
\end{equation}
where $h_{\bx}=\hbar \omega (a\dg_\bx a_\bx + \frac{1}{2})$, 
with $a_\bx=\sqrt{m \omega / (2 \hbar)}  \hat{\bx} + i \sqrt{1/(2m\hbar \omega)} \hat{\bp}_\bx$. 
The wave function of the coherent state $\ket{\alpha_x}$, defined as $a_x \ket{\alpha_x} = \alpha_x \ket{\alpha_x}$, obeys
\begin{equation}
\begin{split}
\alpha_x \braket{x}{\alpha_x} 
=&  \sqrt{\frac{m \omega}{ 2 \hbar}}   x \braket{x}{\alpha_x} + i \sqrt{\frac{1}{2m\hbar \omega}} \frac{\hbar}{i} \frac{\partial}{\partial x} \braket{x}{\alpha_x}.
\end{split}
\end{equation} 
The solution of this equation reads 
\begin{equation}
\braket{x}{\alpha_x} = e^{- \frac{m\omega}{2 \hbar} (x - x_{\alpha_x})^2} e^{i k_{\alpha_x} (x - x_{\alpha_x}) },
\end{equation}
where $(x_{\alpha_x}, k_{\alpha_x})$ are related to the real and imaginary parts of $\alpha_x$ through 
\begin{equation}
\alpha_x = (m \omega x_{\alpha_x} + i \hbar k_{\alpha_x}) / \sqrt{2m\hbar \omega}.
\end{equation}
The 3D wave function of the coherent state for a harmonic oscillator thus reads  
\begin{eqnarray}
\braket{\br}{\ba} =& \braket{x}{\alpha_x} \braket{y}{\alpha_y} \braket{z}{\alpha_z} \nonumber \\
=& e^{-\frac{m \omega}{2\hbar} (\br-\br_\alpha)^2}e^{i \bk_\alpha \cdot (\br - \br_\alpha)},
\end{eqnarray}
where $\br_\alpha = (x_{\alpha_x} {\bf x} + y_{\alpha_y} {\bf y} + z_{\alpha_z} {\bf z})$
and $\bk_\alpha = (k_{\alpha_x} {\bf x} + k_{\alpha_y} {\bf y} + k_{\alpha_z} {\bf z})$.

If we now compare this wave function with the one of the $\ket{\bR,\bK,T}$ state defined in Eqs. (\ref{eq:rwf},\ref{eq:RKTt}), namely,  
\begin{eqnarray}
\braket{\br}{\bR,\bK,T}  & =& L^{-D/2} e^{i \bK \cdot (\br - \bR) } \bra{\br-\bR}e^{-\beta H_0/2} \ket{\br = \bold{0}}\nonumber \\
&=&L^{-D/2} e^{-\frac{2 |\br- \bR|^2}{\lambda_T^2}} e^{i \bK \cdot (\br- \bR)},
\end{eqnarray}
we recognize the coordinate representation of a 3D coherent state by identifying,
\begin{equation}
\begin{split}
\ba =&  \left( m \frac{ k_B T}{\hbar} \bR + i \hbar \bK \right)/ \sqrt{2m k_B T}\\
=& \frac{\bR}{\lambda_T} + i \frac{\lambda_T}{\sqrt{2}} \bK.
\end{split}
\end{equation}
Hence the $\ket{\bR, \bK,T}$ states can be interpreted as the eigenstates of the 3D harmonic oscillator 
with frequency $\hbar \omega = k_B T$. 
%
Coherent states form a well-known basis that has found numerous physical applications. 
It is very likely that  our $\ket{\bR,\bK,T}$ states also find various applications.

\section{Discussion}\label{sec:discussion}

We first wish to note that the splitting of $T$ into $T_\mathcal{R} + T_\mathcal{K}$, as done in Eq.~(\ref{eq:splitting}), appears in a natural way when considering correlation functions. Indeed, the correlation of two operators, $A$ and $B$, namely,  $\langle A(t) B(0) \rangle = {\rm Tr} (A(t) B(0) \rho_{\rm th})$, can be obtained from the thermal density matrix  $\rho_{\rm th} = e^{-\beta H}/{\rm Tr (e^{-\beta H})}$, which corresponds to the normalized Boltzmann operator in the subspace of interest. The Fourier transform of the correlation function yields the retarded Green's function, which for one free particle  reads as (see, e.g., Chap. 4 in \cite{Schwabl2005a})
\begin{align} \label{eq:G}
G^>_{AB}(\omega) &\equiv \int dt \, e^{i \omega t} \langle A(t) B(0) \rangle   \\
&=  \int dt \, e^{i \omega t} \, {\rm Tr}\left( e^{ i H_0 t} A e^{-i H_0 t} B \frac{e^{-\beta H_0}}{{\rm Tr (e^{-\beta H_0})}} \mathbb{1}_1\right) \nonumber\\
&= \frac{2 \pi}{Z_T} \sum_{\bk \bk'} e^{-\beta \epsilon_\bk} \delta(\epsilon_\bk -\epsilon_{\bk'}+\omega) \bra{\bk}A\ket{\bk'} \bra{\bk'}B\ket{\bk}, \nonumber
\end{align}
where $Z_T = {\rm Tr}(e^{-\beta H_0})$. This expression follows from the conventional representation of the thermal state, that is, the eigenstate representation given in Eq. (\ref{eq:th_ak}), here written in the 1-particle subspace. 
The same function can be evaluated using the presently developed thermally excited wave packets. This is done by using the Boltzmann operator as written in Eq. (\ref{eq:th_aRKT}) and by inserting $\ket{\bk}$-state closure relations. This yields (see Appendix C)
\begin{align}\label{eq:newG}
G^>_{AB}(\omega) =&  \frac{2 \pi}{Z_T} \sum_{\bk \bk'} \delta(\epsilon_\bk -\epsilon_{\bk'}+\omega)  \bra{\bk}A\ket{\bk'} \bra{\bk'}B\ket{\bk}\\
&
\times \left( \frac{\sqrt{\pi}\lambda_{T_\mathcal{K}} \lambda_{T_\mathcal{R}}}{\lambda_T L}  \right)^D \sum_\bK e^{-\beta_\mathcal{K} \epsilon_\bK} e^{-\beta_\mathcal{R} \epsilon_{\bk-\bK}} . \nonumber
\end{align}
Identification of  this expression with Eq.~(\ref{eq:G}) imposes   
\begin{equation}
e^{-\beta \epsilon_\bk} = \left( \frac{\sqrt{\pi}\lambda_{T_\mathcal{K}} \lambda_{T_\mathcal{R}}}{\lambda_T L} \right)^D \sum_\bK e^{-\beta_\mathcal{K} \epsilon_\bK} e^{-\beta_\mathcal{R} \epsilon_{\bk-\bK}},
\end{equation}
which can be evaluated by turning to  continuous $\bk$. We then find  $\beta^{-1} = \beta_\mathcal{R}^{-1} + \beta_\mathcal{K}^{-1}$, which just corresponds to Eq. (\ref{eq:splitting}).

One important question still remains to be answered: how to choose the appropriate decomposition in $T_\mathcal{K}$ and $T_\mathcal{R}$ for a given temperature $T$. In other words, how much energy should be put into the spread of the wave-packet and how much into the distribution of its momentum? 
One approach to tackle this issue is to relate the spatial extension with an effective state temperature: The more spread in space, the lower the temperature.

Another approach  lies in the search for a connection between quantum mechanics and quantum statistical mechanics. The thermally excited wave-packet representation pins down the two essential features required for thermal states, as argued in a recent work \cite{Drossel2017a}, namely stochasticity and  spatial extension of the particle wave function. In a sense, the maximal extension of the wave-packet defines how much stochasticity lies in the state. Hence, the answer to how  the thermal energy must be split between the spatial extension of the wave-packet and the momentum distribution can lie in the intrinsic stochasticity of the state. Such an interpretation would support the argument that  the stochasticity and irreversibility in statistical mechanics reflects the true features of nature, but this idea needs further investigation. 
%

We wish to also note that, among the current methods that successfully describes many-body systems at finite temperature, molecular dynamics  simulations \cite{Marx2009a} rely on point particles with a finite spatial extension. The spatial extension is defined either by the electronic shell or by the nuclei thermal wavelength obtained from path integral \textit{ab initio} calculations that account for electronic properties of  electrons or nuclei.
The advantage of the wave-packet representation we  propose here is to treat the particle and wave properties of the matter on equal footing. This is of particular importance for fermions, where the Pauli exclusion principle  limits the state occupation in the wave-packet distribution when $N\geq 2$. Quantum features are relevant when the density is large enough such that the wave-packets display some nonzero overlap.

\textbf{In conclusion}, we here propose a general formalism that represents thermally-equilibrated systems of free massive  bosons or fermions at finite temperature, in terms of thermally-excited wave-packets. 
   This formalism contains an intrinsic flexibility in the spatial extension of the wave-packets created by the operator defined in Eq. (\ref{eq:aRKT}), which is chosen by splitting the  energy according to Eq. (\ref{eq:splitting}) \cite{Hornberger2003a}.  
We can then construct a many-body basis composed of wave-packets chosen according to the relevant physical length of the system of interest. 
These wave-packets, which have the features of classical particles,  provide the missing link for a continuous connection between classical and quantum representations of a thermal gas. 

\emph{Acknowledgments.---}  
We are grateful to John Sipe and Agata Bra\'nczyk for the many helpful discussions.   A.C. thanks the \textit{Institut des NanoSciences de Paris} and J. Cao for hosting her, and acknowledges financial support from the Swiss National Science Foundation.


%

\newpage

\appendix

\section{wave-packet operators}

We provide here some additional results.

The norm of the $\ket{\bR, \bK}$ state is given using Eq. (1) by    
\begin{equation} \label{Seq:norm}
\begin{split}
\braket{\bR,\bK}{\bR,\bK} &= \sum_\bk \braket{ \bR,\bK}{\bk} \braket{ \bk}{\bR,\bK} \\
&= \sum_\bk \braket{\phi}{\bk-\bK} \braket{\bk-\bK}{\phi} = \braket{\phi}{\phi}.
\end{split}
\end{equation}

From the definition of the $a\dg_{\bR,\bK}$ operator (\ref{eq:wp_def}), it is easy to show that
\begin{eqnarray}
[a_\bk, a\dg_{\bR,\bK}] = \braket{\bk}{\bR,\bK}, & & [a_\br, a\dg_{\bR,\bK}] = \braket{\br}{\bR,\bK} \label{Seq:comm_r}\, , \\
\left[a_{\bR',\bK'}, a\dg_{\bR,\bK} \right]&=& \braket{\bR',\bK'}{\bR,\bK} . \nonumber
\end{eqnarray}

We now look at the free Hamiltonian in the wave-packet basis, given in Eq. (\ref{eq:H0}).
In the case of a highly peaked  distribution $\braket{\bk}{\phi} = \delta_{\bk \mathbf{0}}$, the $\phi$ part of Eq. (\ref{eq:H0})  reduces to $\delta_{\bK' \bK} \delta_{\bp \mathbf{0}}$. So, $H_0$ appears as  
\begin{equation}
\sum_\bK \epsilon_\bK \int_{L^D} \frac{d\bR}{L^D} \left( \int_{L^D} \frac{d\bR'}{L^D} a\dg_{\bR',\bK} e^{i \bK \cdot (\bR'-\bR)} \right) a^{}_{\bR,\bK}.
\end{equation}
and  the $a\dg_{\bR,\bK}$ operators reduce to $a\dg_\bK e^{- i \bK \cdot \bR} $, which yields  the simple result given in the main text.

\section{State characterization}
We detail below some of the state properties, keeping the same numeration as in Section V for clarity. 
\subsection{$\ket{\bR,T}$ state}
(ii) The momentum mean value of the state $\ket{\bR,T}$ reads, using Eq.~(32) and $\cav a_\br \, \hat{\bk} \, a\dg_{\br'} \vac = \sum _\bk \bk \braket{\br - \bR}{\bk} \braket{\bk}{\br' - \bR}$,   
 \begin{eqnarray}
 \bra{\bR,T} \hat{\bk} \ket{\bR,T}   &=& \sum_\bk \bk \left | \int_{L^D} d\br \braket {\phi_T}{\br-\bR} \braket{\br-\bR}{\bk} \right|^2 \nonumber \\
 &=& \sum_\bk \bk \, \big|\braket{\bk}{\phi_T} \big|^2,  
 \end{eqnarray}
 which reduces to zero for $|\braket{\bk}{\phi_T}|^2 = |\braket{-\bk}{\phi_T}|^2$.

(iii) The energy of the $\ket{\bR, T}$ state follows from Eq. (32) as  
\begin{align}
& \bra{\bR,T} H_0 \ket{\bR,T} =\\
&  \int_{L^D} d\br d\br' \braket{\phi_T}{\br-\bR} \braket{\br'-\bR}{\phi_T} \cav a_\br \, H_0 \, a\dg_{\br'} \vac.  \nonumber
 \end{align}
To decouple $\br$ from $\br'$, we rewrite  $H_0$  as $\sum_\bk \epsilon_\bk \braket{\br-\bR}{\bk} \braket{\bk}{\br'-\bR}$. Integrations over $(\br,\br')$, readily performed through closure relations, give the r.h.s. of the above equation as $\sum_\bk \epsilon_\bk |\braket{\bk }{\phi_T} |^2 =L^{-D} \sum_\bk \epsilon_\bk e^{-\beta \epsilon_\bk}$. 
So, the energy of the $\ket{\bR,T}$ state reduces to  
\begin{equation} 
\frac{\bra{\bR, T} \: H_0 \: \ket{\bR, T} }{\braket{\bR, T}{\bR, T}}  = \frac{\sum_\bk \epsilon_\bk e^{-\beta \epsilon_\bk}}{\sum_\bk e^{-\beta \epsilon_\bk}} = \frac{D}{2} k_B T, 
\end{equation}

(iv) The energy variance of the $a\dg_{\bR, T}\vac$ state is given by  
\begin{equation}
\begin{split}
\sigma_{\bR,T} =&\frac{ \bra{\bR, T} \: H_0^2 \: \ket{\bR T} }{\braket{\bR, T}{\bR ,T}} 
  - \frac{\bra{\bR, T} \: H_0 \: \ket{\bR, T} ^2 }{\braket{\bR, T}  {\bR ,T}^2 }    \\
=& \frac{D}{2} (k_B T)^2.
\end{split}
\end{equation}

\subsection{$\ket{\bR,\bK,T}$ state}

(i) The operator $a\dg_{\bR,\bK,T}$ defined in Eq.~(\ref{eq:aRKT}) creates a state $\ket{\bR, \bK ,T}= a\dg_{\bR, \bK, T} \vac$ having a  wave function 
\begin{equation}
\braket{\br}{\bR, \bK, T} = L^{-D/2} e^{i \bK \cdot(\br-\bR)}\braket{\br-\bR}{\phi_T},
\end{equation}
 similar to $\braket{\br}{\bR,\bK}$ defined in Eq.~(\ref{eq:rwf}) with $\ket{\phi}$ replaced by $L^{-D/2} \ket{\phi_T}$. 
In the same way, Eq.~(\ref{eq:aRKT}) gives 
\begin{equation}
\braket{\bk}{\bR,\bK,T} = L^{-D/2} e^{-i \bk \cdot \bR} \braket{\bk - \bK}{\phi_T},
\end{equation}
which is similar to $\braket{\bk}{\bR,\bK}$ in Eq.~(\ref{eq:kwf}) within the same replacement. 

(iii) The energy  mean value of the $\ket{\bR, \bK, T}$ state follows from 
$ \bra{\bR, \bK ,T} H_0 \ket{\bR, \bK, T} {=} L^{-2D} \sum_{\bk}  \epsilon_\bk |\braket{\bk{-} \bK}{\phi_T}|^2 $, 
as given in Eq.~(\ref{eq:H0}) with $\ket{\phi}$ replaced by $L^{-D/2} \ket{\phi_T}$. 

(iv) Similarly, we can derive the variance in energy. From  
$\bra{\bR, \bK, T} H_0^2 \ket{\bR, \bK, T}  =L^{-2D} \sum_\bk \epsilon_{\bk+ \bK}^2 e^{-\beta \epsilon_\bk}$,  
we find that the variance given in (\ref{eq:sigma_wp}) barely follows from its definition 
\begin{eqnarray*} \label{Seq:sigma_wp}
\sigma_{\bR, \bK, T} &=&\frac{ \bra{\bR, \bK, T} \: H_0^2 \: \ket{\bR, \bK, T} }{\braket{\bR,\bK,T}{\bR,\bK,T}}
  -  \frac{\bra{\bR, \bK, T} \: H_0 \: \ket{\bR, \bK, T}^2  }{\braket{\bR,\bK,T}{\bR,\bK,T}^2}. 
\end{eqnarray*}

(v) The position expectation value for the time-dependent wave-packet, defined in Eq. (\ref{eq:evolution}), is obtained by splitting $\bra{\bk}\hat{\br}\ket{\bk'}$ through the $\ket{\br}$-state closure relation,  
\begin{align}
&{_t \bra{\bR,\bK,T} \hat{\br}\ket{\bR,\bK,T}_t} \\
&=\  \sum_{\bk\bk'} {}_t\braket{\bR,\bK,T}{\bk} \bra{\bk}\hat{\br}\ket{\bk'} \braket{\bk'}{\bR,\bK,T}_t \nonumber \\
&= L^{-D}\int_{L^D} d\br \, \br \left| \sum_\bk \braket{\phi_T}{\bk-\bK} \braket{\bk}{\br} e^{i (\bk\cdot \bR+ \epsilon_\bk t)} \right|^2. \quad \qquad \nonumber
\end{align}
To go further, we note that $\braket{\phi_T}{\bk-\bK} $ forces $\bk \approx \bK$. This leads us to expand $\epsilon_\bk$ as $\approx \epsilon_\bK + (\bk-\bK) \cdot  \bK /m$ for $\hbar=1$. So, by setting  $ \bR +  \bK t /m = \bR_t$, we find 
\begin{eqnarray}
&&{_t \bra{\bR,\bK,T} \hat{\br}\ket{\bR,\bK,T}_t} \nonumber\\
&=& \frac{1}{L^D}\int_{L^D} d\br \: \br \left|\sum_\bk  \braket{\phi_T}{\bk-\bK}\braket{\bk}{\br} e^{i (\bk- \bK)\cdot \bR_t} \right|^2 \nonumber \\
&\approx&\frac{1}{L^D} \int_{L^D} d\br \: \br |\braket{\phi_T}{\br-\bR_t}|^2 =\frac{ \braket{\phi_T}{\phi_T}}{L^D}  \bR_t , \quad \qquad \qquad
\end{eqnarray}

\section{Retarded Green's function}
We provide here details for the calculation of the retarded Green's function written in the wave-packet basis (\ref{eq:newG}). 
Starting from its definition (\ref{eq:G}) and using the Boltzmann operator represented in the wave-packet basis (\ref{eq:th_aRKT}), we  insert $\ket{\bk}$-state closure relations to get 
\begin{align}
&G^>_{AB}(\omega) =  \frac{1}{Z_T} \Big(\frac{\sqrt{\pi}\lambda_{T_\mathcal{R}}\lambda_{T_\mathcal{K}}}{\lambda_{T}}\Big) 
\sum_{\bk,\bk',\bk''}\int dt \, e^{i (\omega+\epsilon_\bk - \epsilon_{\bk'}) t}  \nonumber  \\
&\times \sum_\bK e^{-\beta_\mathcal{K} \epsilon_\bK}  \int_{L^D}d\bR  \bra{\bk}A\ket{\bk'} \bra{\bk'}B\ket{\bk''}\braket{\bk''}{\bR,\bK,T_\mathcal{R}}\braket{\bR,\bK,T_\mathcal{R}}{\bk} . \nonumber
\end{align}
The Green's function given in Eq. (\ref{eq:newG}) then follows from  
\begin{equation}
\int_{L^D} d\bR \braket{\bk''}{\bR,\bK,T} \braket{\bR, \bK,T}{\bk} = \delta_{\bk'' \bk} |\braket{\bk-\bK}{\phi_T}|^2,
\end{equation}
and the $\bk$ representation of the $\ket{\phi_T}$ state  
\begin{eqnarray}
\braket{\bk}{\phi_T} &=& \bra{\bk} e^{-\beta H_0/2} \ket{\br=0 } \nonumber \\
&=& e^{-\beta \epsilon_\bk /2} L^{-D/2}. 
\end{eqnarray}

\vfill

\end{document}